# Estimation of Lateral Distribution Function in Extensive Air Showers by Using AIRES Simulation System

Al-Rubaiee A. Ahmed, Ahmed Jumaah
Al-Mustansiriyah University, College of Science, Department of Physics

## Abstract

In this work the estimation of the lateral distribution function in Extensive Air showers was performed by using a system for air shower simulations which is called AIRES version 2.6 for different hadronic models like (QGSJET99, SIBYLL and SIBYLL1.6). The simulation was fulfilled in the high energy range ($10^{15}$-$10^{19}$ eV) for different primary particles like (gamma, protons and iron nuclei) for vertical showers. This simulation can be used to reconstruct the type and energy of the particle that generated Extensive Air showers for charged particles that registered with different arrays.

## 1-Introduction

High energy cosmic rays can be studied depending on experimental devices that located on the surface of the Earth. This means that such cosmic rays cannot be detected directly; it is necessary instead to measure the products of the atmospheric cascades of particles initiated by the incident astroparticle. When the primary cosmic particle interacts with the Earth's atmosphere, the atmospheric particle shower will begins directly. In general, this is an inelastic nuclear collision that generates a number of secondary particles [1-3]. Those particles continue interacting and generating more secondary particles which in turn interact again similarly as their predecessors. This multiplication process continues until a maximum is reached. Then the shower attenuates as far as more and more particles fall below the threshold for further particle production [4]. A detailed information of the physics involved is thus necessary to explain appropriately the measured observables and be able to infer the properties of the primary particles [1].

This problem can be involved by many aspects: Interactions of high energy particles, properties of the atmosphere and the geomagnetic field, etc. Computer simulation is one of the most appropriate tools to quantitatively analyze such particle showers. The AIRES system is a set of programs to simulate such extensive air showers (EAS) [1]. One of the basic objectives considered during the development of the software is that of designing the program modularly, in order to make it easier to switch among the different models that are available, without having to get attached to a particular one [5, 6].

In this work the simulation of the EAS was performed by estimating the lateral distribution function (LDF) by using the AIRES code for different hadronic models using (QGSJET99 [7]; SIBYLL [8] and SIBYLL S16 [9]) packages for the simulation of hadronic processes. The simulation was performed for different primary particles like (Gamma, Protons and Iron nuclei) at the energy range ($10^{15}$-$10^{19}$) for vertical showers.

## 2-AIRES Simulation System

The name of AIRES [5, 6] (AIR-shower Extended Simulation) identifies a set of programs and subroutines to simulate particle showers produced after the incidence of high energy cosmic rays on the Earth atmosphere, and to manage all the related output data. AIRES provides full space-time particle propagation in a realistic environment where the characteristics of the atmosphere, the geomagnetic field and the Earth curvature are taken into account adequately.

The particles taken into account by AIRES in the simulations are: Gammas, electrons, positrons, muons, pions, mesons, lambda baryons, nucleons, antinucleons and nuclei up to Z = 36. Electron, muon and neutrinos are generated in certain processes (decays) and accounted for their energy but not propagated. The primary particle can be any one of the already mentioned particles, with energy ranging from less than $10^9$ eV up to more than $10^{21}$ eV.

AIRES system has been successfully used to study several characteristics of the showers, including comparisons between interaction models and/or experimental data. We have used AIRES system to estimate the influence of diffractive processes in the final shower observables, and to study in detail the characteristics of showers initiated by photons in connection with cosmic ray composition analysis at the highest energies.

## 3- Results and Discussion:

The simulation of the LDF in EAS was obtained by using AIRES code for primary particles (Gamma, protons and iron nuclei) at the highest energies range from ($10^{15}$-$10^{19}$ eV) for vertical showers.

The degree of reduction of the fluctuation does depend on the observable considered. In figures (1-3) the lateral distribution of ground gamma, protons and iron nuclei is displayed, again for different thinning levels. It is noticeable the degree of persistence of the noisy fluctuations, which are not completely eliminated even in the $10^{-7}$ relative thinning case.

Figure (1) shows the LDF for gamma particle from the core to radial distances by AIRES simulation and using hadronic models (QGSJET99, SIBYLL16 and SIBYLL) with the energy $5.0\times10^{17}$ eV. From the figure, one can observe that the LDF is directly proportional with the primary energy of the particle i.e. when the energy of primary particle increases the LDF increases too.

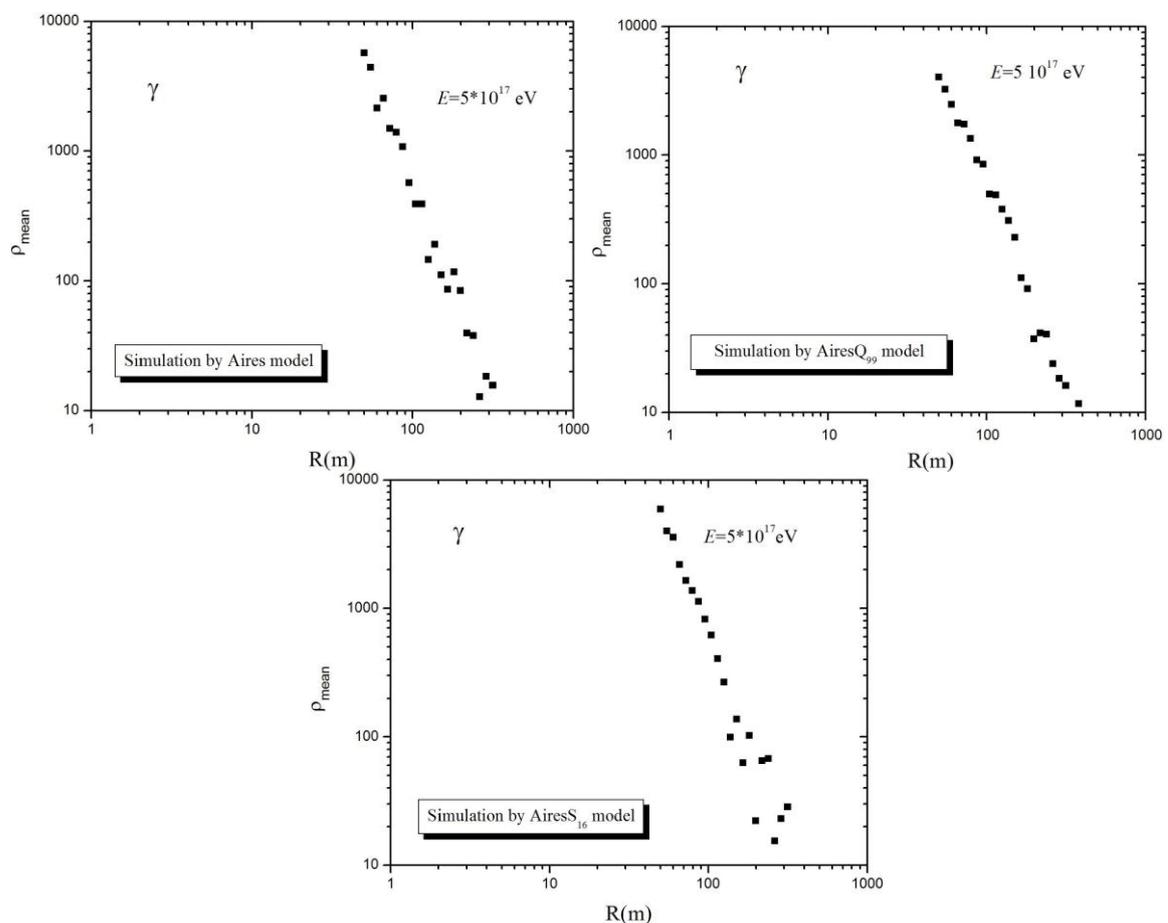

**Figure 1** The lateral distribution function simulated by AIRES code for vertical shower at the energy ($5*10^{17}$ eV) for gamma particle

The lateral distribution of primary protons displayed in figure (2) reflects another characteristic of the thinning algorithm. Even if the fluctuations are very large for $10^{-4}$ relative thinning level, they reduce immediately when the thinning is lowered. To understand the behavior of these distributions it is necessary to recall that the protons undergo a much reduced number of interactions before reaching ground.

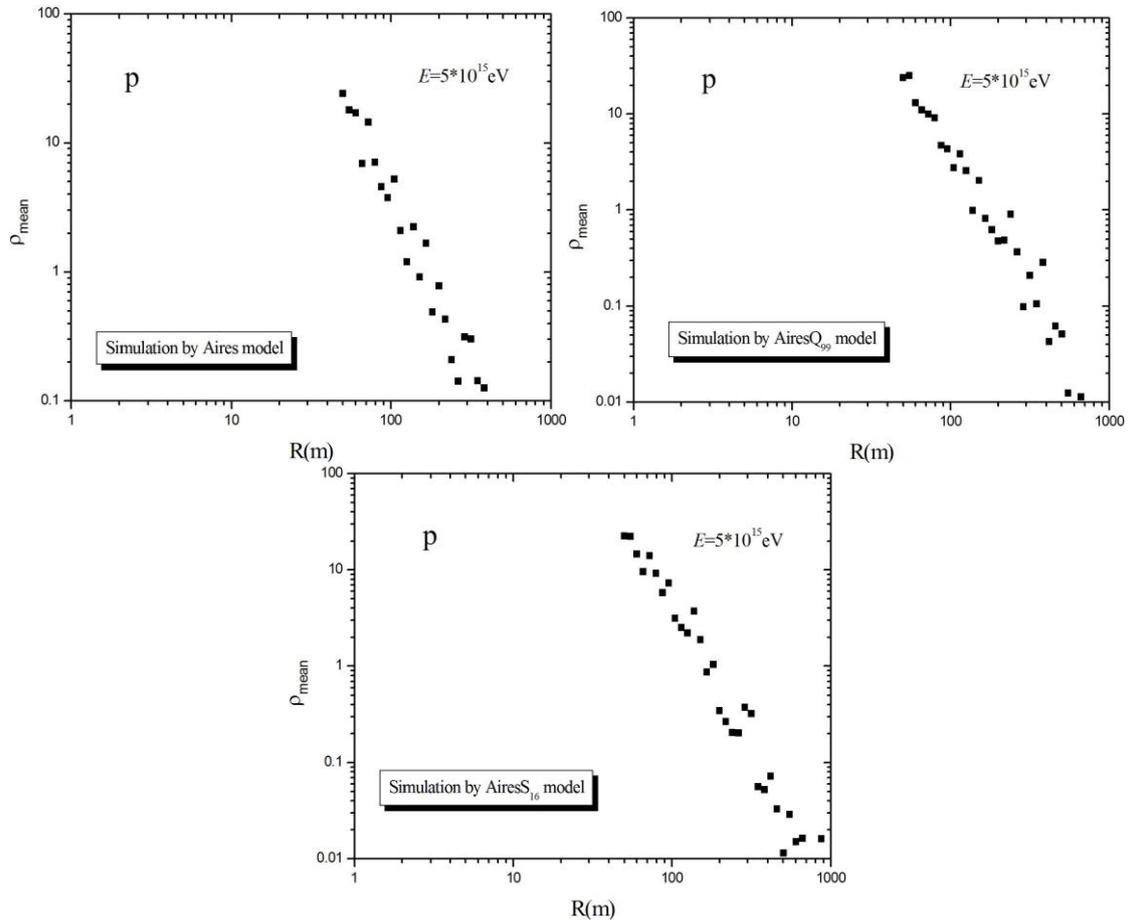

**Figure 2** The lateral distribution function simulated by AIRES code for vertical showers at the energy ($5\times10^{15}$ eV) for primary proton.

The figure (3) shows the results of simulated LDF in vertical EAS initiated by primary iron nuclei for energy ($5\times10^{19}$ eV). Through this figure one can see the effect of the thinning energy on the fluctuations of the lateral distribution of iron nuclei, in the same conditions as in figures 2 and 3 for different hadronic models.

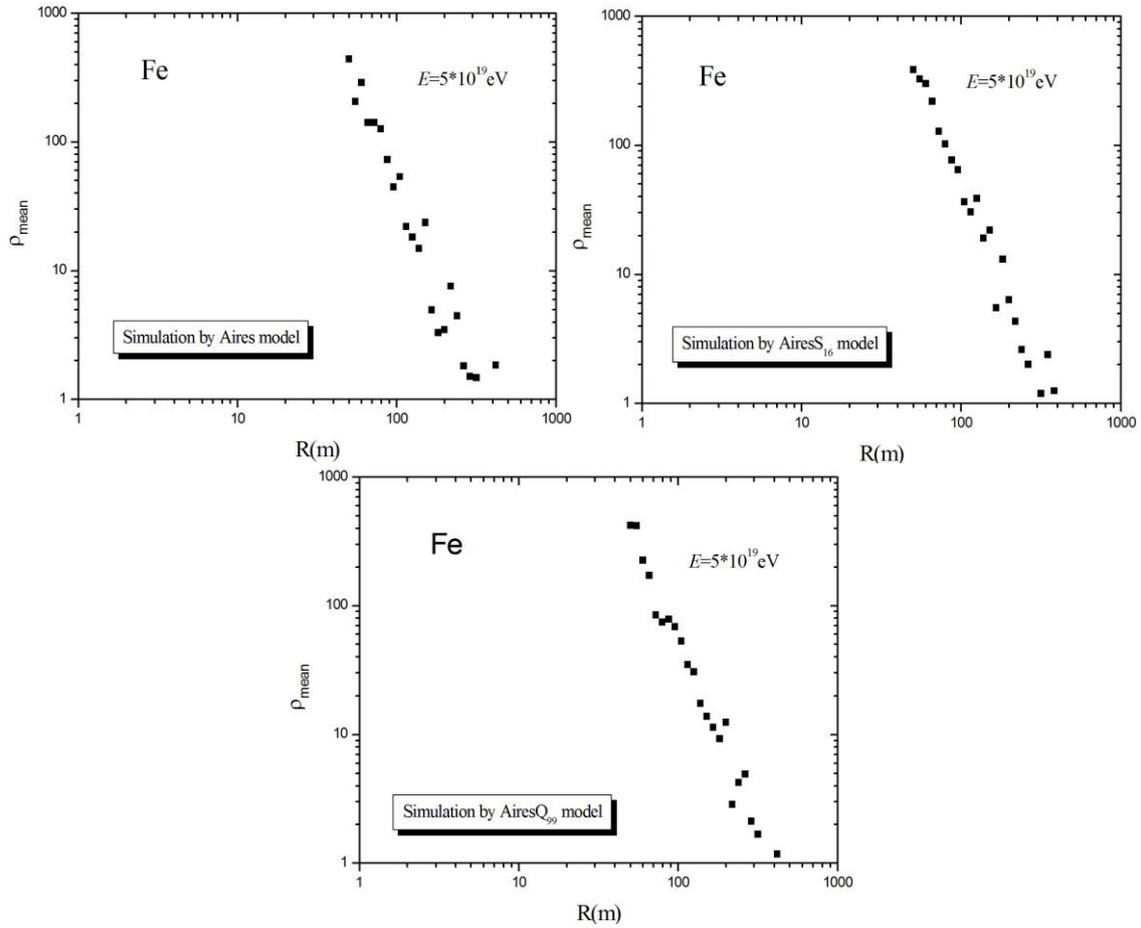
**Figure 3** The lateral distribution function simulated by AIRES code for vertical shower at the energy ($5\times10^{19}$ eV) for iron nuclei

## Conclusion

In the present work the simulation of lateral distribution function was simulated by using AIRES system for three different models like (QGSJET99; SIBYLL and SIBYLL S16) for different primary particles (gamma, protons and iron nuclei) at the energy range ($10^{15}$-$10^{19}$ eV). The simulation of the lateral distribution function has shown an opportunity of primary particle identification and definition of its energy around the knee of the cosmic ray spectrum. The main advantage of the given approach consists of the possibility to make a library of lateral distribution function samples which could be utilized for analysis of real events which detected with the extensive air shower array and reconstruction of the primary cosmic rays energy spectrum and mass composition.